\documentclass[aps,prl,twocolumn,groupedaddress,superscriptaddress,showpacs]{revtex4}
\usepackage{natbib}
\usepackage[latin1]{inputenc}
\usepackage[dvips]{graphicx}
\usepackage{amsmath,amsbsy,amsfonts,amssymb}
\usepackage{bm}

\newcommand{\abs}[1]{\left\vert#1\right\vert}
\newcommand{\state}[3]{\ensuremath{\,^{#1}{#2}_{#3}}}
\newcommand{\unit}[1]{\ensuremath{\,\mathrm{{#1}}}}

\begin{document}
\title{Experimental investigation of ultracold atom-molecule collisions}
\author{Peter Staanum}
\affiliation{Physikalisches Institut, Universit\"at Freiburg,
Herman-Herder-Strasse 3, 79104 Freiburg, Germany.}
\affiliation{Institute of Quantum Optics, Universit\"at Hannover,
Welfengarten 1, 30167 Hannover, Germany.}
\date{\today}
\author{Stephan D. Kraft}
\affiliation{Physikalisches Institut, Universit\"at Freiburg,
Herman-Herder-Strasse 3, 79104 Freiburg, Germany.}
\author{J\"org Lange}
\affiliation{Physikalisches Institut, Universit\"at Freiburg,
Herman-Herder-Strasse 3, 79104 Freiburg, Germany.}
\author{Roland Wester}
\affiliation{Physikalisches Institut, Universit\"at Freiburg,
Herman-Herder-Strasse 3, 79104 Freiburg, Germany.}
\author{Matthias Weidem\"uller}
\affiliation{Physikalisches Institut, Universit\"at Freiburg,
Herman-Herder-Strasse 3, 79104 Freiburg, Germany.}
\date{\today}

\begin{abstract}
Ultracold collisions between Cs atoms and $\rm{Cs}_2$ dimers in
the electronic ground state are observed in an optically trapped
gas of atoms and molecules. The $\rm{Cs}_2$ molecules are formed
in the triplet ground state by cw-photoassociation through the
outer well of the $0_g^-(P_{3/2})$ excited electronic state.
Inelastic atom-molecule collisions converting internal excitation
into kinetic energy lead to a loss of $\rm{Cs}_2$ molecules from
the dipole trap. Rate coefficients are determined for collisions
involving Cs atoms in either the F=3 or F=4 hyperfine ground state
and $\rm{Cs}_2$ molecules in either highly vibrationally excited
states ($v'=32-47$) or in low vibrational states ($v'=4-6$) of the
$a ^3\Sigma_u^+$ triplet ground state. The rate coefficients
$\beta\sim10^{-10}$\,cm$^3$/s are found to be largely independent
of the vibrational and rotational excitation indicating unitary
limited cross sections.
\end{abstract}

\maketitle


At very low temperatures the elastic and inelastic scattering of
particles is governed by quantum mechanical effects like
resonances or tunneling through barriers, since the deBroglie
wavelength becomes comparable to the range of the interparticle
interactions. With the advent of laser cooling techniques,
ultracold atom-atom collisions have been studied intensively both
experimentally and theoretically. Their understanding and control
has been decisive for the rapid development of the field of
quantum gases~\citep{Weiner-CollisionsReview}. As a prominent
example, the competition between elastic and inelastic collisions
of trapped atoms determines the success of evaporative cooling for
achieving atomic Bose-Einstein condensation~\citep{Anderson-BEC}.
Since recently, the focus has been moving from pure atomic to
ultracold atom-molecule and molecule-molecule collisions, which
are highly relevant for the formation and stability of molecular
Bose-Einstein condensates~\citep{jochim2003:sci,greiner2003:nat}.
The investigation of atom-molecule scattering at very low
temperatures forms the basis of the new emerging field Ultracold
Chemistry~\citep{Krems-UltracoldReview}.


Theoretical studies show that ultracold atom-molecule collisions
do not ``freeze-out'' at low temperatures, as classical Langevin
capture theory predicts, but feature instead significant rate
coefficients at ultralow temperatures where only a few partial
waves
contribute~\citep{Balakrishnan-HeH2,Balakrishnan-HH2collisions,Cvitas-LiLi2,Quemener-KK2,Quemener-NaNa2}.
The energy dependence of cross sections at low temperatures is
predicted to be in good agreement with the Wigner threshold
law~\citep{Cvitas-LiLi2,Quemener-KK2,Quemener-NaNa2}. Up to now
these calculations are based on single potential energy surfaces.
However, coupling of several low lying potentials through conical
intersections can lead to additional features in the rate
coefficients through shape and Feshbach
resonances~\citep{brue-li-potential}. Since the collision energy
is much smaller than the energy spacing between internal states in
the colliding atoms and molecules, the only inelastic processes to
occur are collisional deexcitation and, for non-identical atoms,
reactive scattering followed by chemical rearrangement. At room
temperature, the energy transfer from molecular vibration to
translation and rotation is suppressed when the vibrational level
spacings are large compared to the translational temperature,
i.e., the probability for deexcitation per collision is much
smaller than unity~\citep{SSH-theory,wodtke-review}. In contrast,
below $10^{-3}$\,K Refs.~\citep{Cvitas-LiLi2, Quemener-NaNa2,
Quemener-KK2} show that the inelastic rate coefficient for
lithium, sodium and potassium atom-diatom collisions is higher
than the elastic rate coefficient and does not depend on
temperature, again in agreement with the Wigner threshold law. In
Refs.~\citep{Balakrishnan-HeH2,Balakrishnan-HH2collisions} the
inelastic collision rate coefficient for He-H$_2$ and H-H$_2$
collisions is found to strongly increase with the vibrational
quantum number. On the other hand
Stwalley~\citep{Stwalley-CanJPhys} has recently argued that
molecules in the highest vibrational levels near the dissociation
threshold should feature very small inelastic collision rates. In
addition to vibrational excitation, rotational quantum states may
influence the rate of inelastic collisions, e.g., due to
quasi-resonant exchange of rotational and vibrational
energy~\citep{Forrey1999}. Clearly, experimental investigations
are needed to reveal the collisional behaviour of molecules at
ultralow temperatures and to test quantum scattering calculations
in the interesting transition region from the Wigner threshold
regime to the Langevin regime.

Few experimental studies on ultracold atom-molecule collisions
have been carried out so far, the possibilities being limited by
the difficulty to produce and trap ultracold molecules
simultaneously with cold atoms at sufficiently high densities.
Successful storage of ultracold molecules has first been shown in
an optical dipole trap~\citep{Takekoshi-OpticalTrappingCs2} and
recently in magnetic traps~\citep{Vanhaecke-MagneticTrap}. Cold
collisions at a temperature of $T\sim0.5\unit{K}$ have been
studied between He buffer gas cooled CaF and
He~\citep{Maussang-CaFHe}. For ultracold $\rm{Na}_2$ molecules
formed through a Feshbach resonance, collision rate coefficients
were determined for both atom-molecule and molecule-molecule
collisions~\citep{Mukaiyama-CollNaMol}. Similarly ultracold
molecule-molecule collisions were observed for $\rm{Cs}_2$ and a
strong loss rate due to Feshbach-like resonances was
observed~\citep{Chin-MolColl}. These two experiments at ultralow
temperatures probe molecules in the last bound vibrational level.


Here we present an experimental study of ultracold atom-molecule
collisions involving molecules in deeply bound vibrational states.
Cs ground state atoms and photoassociated $\rm{Cs}_2$ molecules
are stored together in an optical dipole trap. Rate coefficients
are determined for exoergic atom-molecule collisions, observed
through molecule loss, with Cs$_2$ in high lying vibrational
levels ($v'=32-47$, bound by 1-24\unit{cm^{-1}}) and in low lying
levels ($v'=4-6$, bound by 175-195\unit{cm^{-1}}) in the $a
^3\Sigma^+$ triplet ground state~\citep{Olivier-private-v6,
Olivier-private-v79}. We find for molecules in high lying and low
lying states equally large rate coefficients, larger than the
s-wave scattering limit. The inelastic rate coefficients for
different Cs$_2$ rotational states coincide, which indicates that
vibrational deexcitation dominates over rotational deexcitation in
the collision. A significant increase in the deexcitation rate is
found for collisions with Cs atoms in the F=4 hyperfine level of
the \state{2}{S}{1/2} ground state as compared to Cs(F=3).


The measurements proceed as follows. An optical dipole trap,
formed by the focus of a 120\unit{W} $\rm{CO}_2$ laser (Synrad
Evolution 100, waist $\sim 90\unit{\mu m}$) is loaded with Cs(F=3)
atoms from a magneto-optical trap (MOT). We transfer $N_{at}\sim
2\times 10^5$ Cs atoms into the dipole trap with a shot-to-shot
variation of less than 5\% at a typical mean density of
$n_{at}=10^{11}\unit{cm^{-3}}$ and temperature $T=60\pm20\unit{\mu
K}$. The atom density is determined and the temperature estimated
from a measurement of the F=4 atom loss due to hyperfine-changing
collisions~\citep{Mudrich-collisions} and atom-background gas
collisions. The atoms are then optically pumped to F=4 and about
5\% of the atoms are transformed into molecules by
photoassociation~\citep{Wester-PAinDipoleTrap} with a cw
Ti:Sapphire laser (Coherent MBR110) applied with intensity $\sim
600\unit{W/cm^2}$. After 100\unit{ms} of photoassociation an
equilibrium between formation and collision induced loss has been
reached, resulting in $\sim 5000$ molecules in the trap. Their
temperature is estimated to be approximately equal to the atom
temperature, i.e., 60\unit{\mu K}, and the mean density
$n_{mol}\sim 3\times 10^9\unit{cm^{-3}}$. Subsequently, the
remaining atoms can be either optically pumped to the F=3
hyperfine state ($>99\%$ efficiency with negligible atom loss), be
pushed out of the trap with a resonant laser beam ($>99\%$
efficiency) or be left in the F=4 state. By measuring the number
of trapped molecules $N_{mol}$ as a function of storage time
$\tau$ we deduce the inelastic collision rate coefficient. Each
measurement of $N_{mol}(\tau)$ is an average over typically 48
cycles. $N_{mol}(\tau=0)$ is determined before and after each
measurement to correct for drifts in the experiment, mainly small
frequency drifts of the photoassocation laser. The error bars of
the detected molecule number combine the corresponding statistical
and correction uncertainties.

\begin{figure}[!tbp]
\centering\includegraphics[width=0.8\linewidth]{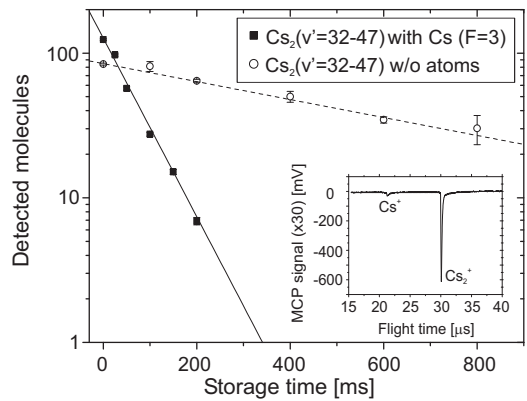}
\caption{Storage time measurement for $\rm{Cs}_2$$(v'=32-47)$
molecules with Cs(F=3) atoms in the dipole trap at a mean density
of
$n_{at}=9.1\pm0.4_{\rm{stat}}\pm2.4_{\rm{sys}}\times10^{10}\unit{cm^{-3}}$
(black filled squares) and with only molecules in the trap
(circles). The lines represent fits to an exponential decay.
Inset: Time-of-flight spectrum showing the $\rm{Cs}^+$ and
$\rm{Cs}_2^+$ signals.}\label{fig1:CollEvidence}
\end{figure}

The molecules are detected via resonance-enhanced two-photon
ionization using an unfocused pulsed dye laser beam (Radiant Dyes
Narrowscan, $\sim10\unit{mJ}$ per pulse, wavelength 712.5\unit{nm}
and linewidth$<0.1\unit{cm^{-1}}$). The $\rm{Cs}_2^+$ ions are
measured with a high resolution time-of-flight mass
spectrometer~\citep{Kraft-TOF}. The number of detected molecules,
calibrated from the signal of individual $\rm{Cs}_2^+$ ions, is
smaller than the number of trapped molecules due to the combined
efficiencies of the ionization (10\%) and detection
(20\%)~\citep{Fraser-MCPefficiency}. This calibration does not
influence the measured rate coefficients, which only depend on the
relative intensity of the molecular signal. Ground state atoms can
only be ionized by three-photon ionization. Therefore, the
$\rm{Cs}^+$ signal is suppressed as compared to the $\rm{Cs}_2^+$
signal (see inset of Fig.~\ref{fig1:CollEvidence}).

The preparation of Cs$_2$ molecules in either high or low lying
levels of the triplet ground state is performed by
photoassociation through the $v=6$ or $v=79$ states, respectively,
of the outer well of the $0_g^-(\state{2}{P}{3/2})$
potential~\citep{Fioretti-Cs20g-Spectroscopy}. The $v=6$ state
decays primarily to the states $v'=32-47$, whereas for $v=79$
mainly the low lying states $v'=4-6$ are
populated~\citep{Olivier-private-v6,Olivier-private-v79}. In the
$0_g^-(v=6)$ state the rotational states are resolved and hence
molecules can be formed in the $a^3\Sigma_u^+$ state in
$(v',J'=1)$ and $(v',J'=J,J\pm 1)$ when $J=0$ and $J\neq 0$,
respectively ($\abs{J-J'}=0,1$; $J=0\nrightarrow J'=0$). In
Fig.~\ref{fig1:CollEvidence} we show examples of storage time
measurements of $\rm{Cs}_2$$(v'=32-47)$ molecules with and without
atoms in the trap. In the presence of atoms at a mean density of
$9.1\pm0.4_{\rm{stat}}\pm2.4_{\rm{sys}}\times10^{10}\unit{cm^{-3}}$
the molecule storage time is reduced by a factor of 7 as compared
to the case without atoms, a clear indication of collision induced
loss.

The collision induced molecule loss can be due to vibrational and
rotational deexcitation of the molecules as well as a change of
the molecular or atomic hyperfine ground state (for Cs in F=4). In
all cases the released energy largely exceeds the trap depth of
$\sim 2\unit{mK}$ for molecules and $\sim1\unit{mK}$ for atoms and
hence both collision partners are lost from the trap with
practically unity probability. Since the number of atoms $N_{at}$
is much larger than $N_{mol}$, the evolution of $N_{mol}$ can be
described by a simple rate equation
\begin{equation}\label{MolDiffEq}
  \dot{N}_{mol}=-\left[\frac{8}{\sqrt{27}}\beta n_{at}(t)+\Gamma_{mol}\right]N_{mol},
\end{equation}
where $\beta$ is the atom-molecule inelastic collision rate
coefficient, $\Gamma_{mol}$ is the loss rate due to
molecule-background gas collisions and $n_{at}$ is weakly
time-dependent due to hyperfine-changing collisions (for Cs(F=4))
and collisions of atoms with background gas. The factor of
$8/\sqrt{27}$ accounts for the spatial averaging over atomic and
molecular spatial distributions assuming equal atom and molecule
temperature. This equation is strictly valid only for a single
internal quantum state of the molecule, due to the general
dependence of the rate coefficient on the quantum state. Therefore
we effectively obtain rate coefficients averaged over the
populated quantum states. Molecule-molecule collisions are
neglected in Eq.~\eqref{MolDiffEq}, because no significant
molecule-molecule collison rate was observed when measuring the
dependence of the molecular storage time on the initial number of
molecules.

\begin{figure}[!tbp]
\centering\includegraphics[width=0.8\linewidth]{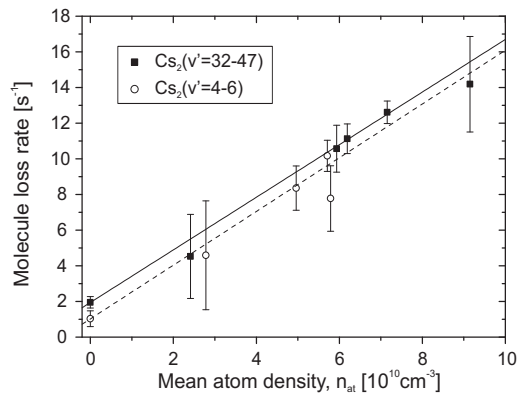}
\caption{Molecule loss rate $8\beta n_{at}/\sqrt{27}+\Gamma_{mol}$
vs. $n_{at}$ for collisions between Cs(F=3) and
$\rm{Cs}_2$$(v'=32-47)$ (solid squares) and $\rm{Cs}_2$$(v'=4-6)$
(circles). The straight lines represent fits discussed in the text
(solid: $v'=32-47$, dashed: $v'=4-6$).}\label{fig:v6v79}
\end{figure}

In the case of F=3 atoms, $n_{at}$ is constant since changes due
to atom-background gas collisions (storage time $\sim 5\unit{s}$)
can be neglected over the relevant molecule storage times. Thus,
$N_{mol}(t)$ is described by an exponential decay with loss rate
$8\beta n_{at}/\sqrt{27}+\Gamma_{mol}$. In Fig.~\ref{fig:v6v79},
this loss rate is shown for $\rm{Cs}_2$$(v'=32-47)$ as well as
$\rm{Cs}_2$$(v'=4-6)$ at different Cs(F=3) atom densities. The
density is varied by varying the MOT size before transfer into the
dipole trap and the final detuning for cooling in optical
molasses~\citep{Wester-PAinDipoleTrap}. For $n_{at}=0$ all atoms
are pushed out of the trap after photoassociation. The loss rate
follows the expected linear dependence on atom density, thus
confirming molecule loss due to atom-molecule collisions. From
fitting a straight line the inelastic rate coefficients
$\beta_{v'=32-47}^{F=3}=0.96\pm 0.02_{\rm{stat}}\pm
0.3_{\rm{sys}}\times 10^{-10}\unit{cm^3/s}$ and
$\beta_{v'=4-6}^{F=3}=0.98\pm 0.07_{\rm{stat}}\pm
0.3_{\rm{sys}}\times 10^{-10}\unit{cm^3/s}$ are deduced (the
systematic error is due to the uncertainty on $n_{at}$ of which
the ratio between collision rate coefficients is independent).
Interestingly, these rate coefficients show no dependence on the
Cs$_2$ vibrational level, even though they bridge many vibrational
quantum numbers. This effect is further discussed below.

In order to investigate to which extent rotational deexcitation
contributes to the molecule loss, we performed molecule storage
time measurements for photoassociation via the rotational levels
$0_g^-(v=6,J=0,1,2,3,4)$ thus populating different rotational
levels $a^3\Sigma_u^+(v'=32-47,J')$. If rotational deexcitation
gave rise to a significant molecule loss we would expect a
relatively low loss rate when $a^3\Sigma_u^+(v'=32-47,J'=0)$ is
populated, i.e., for photoassociation to $J=1$. However, we
consistently find the rate coefficients for all rotational states
$J'$ to be equal within our experimental error. Thus, we conclude
that the molecule loss is governed by vibrational deexcitation.
Molecular hyperfine changing collision rates are expected to be
similar to atomic ones \cite{Mudrich-collisions} and therefore
about an order of magnitude smaller.

In contrast to the ultracold atom-molecule collisions, the
collisions of the optically trapped molecules with the background
gas at 300\,K do depend on the vibrational state, as seen in the
loss rates $\Gamma_{mol}(v'=32-47)=1.9(2)\unit{s^{-1}}$ and
$\Gamma_{mol}(v'=4-6)=1.0(3)\unit{s^{-1}}$ deduced from the fit.
The faster loss for the higher vibrational levels may be
attributed either to a larger geometric cross section for elastic
collisions due to the larger average bond length or to the
energetically open channel of collision induced dissociation
(binding energy $\sim 10\unit{cm^{-1}}$ for $v'=32-47$ compared to
$\sim 185\unit{cm^{-1}}$ for $v'=4-6$). In either case, this
observation shows that the molecules are indeed selectively formed
in either high lying or low lying vibrational states of the
triplet ground state.

\begin{figure}[!tbp]
\centering\includegraphics[width=0.8\linewidth]{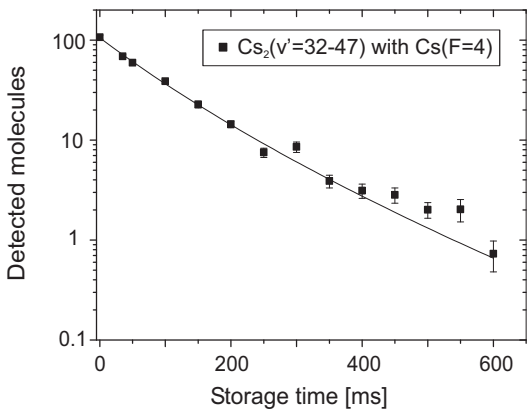}
\caption{Storage time measurement for $\rm{Cs}_2$$(v'=32-47)$ in
collisions with Cs(F=4). Eq.~\eqref{Eq:N2(t),F=4} is fitted to the
data to obtain the inelastic rate coefficient.}
\label{fig5:BetaF4}
\end{figure}

When molecules are stored with Cs(F=4) atoms, a time dependence of
the atomic density has to be included due to hyperfine-changing
atom-atom collisions (see Eq.(7) of
Ref.~\citep{Mudrich-collisions}). Solving Eq.~\eqref{MolDiffEq} we
find
\begin{equation}
N_{mol}(t)=N_{mol}(0)e^{-\Gamma_{mol}t}[1+7Gn_{at}(0)t/2]^{-16\beta/7\sqrt{27}G},\label{Eq:N2(t),F=4}
\end{equation}
where $G=1.1\pm0.3\times 10^{-11}\unit{cm^3/s}$ is the rate
coefficient for hyperfine-changing collisions between Cs(F=4)
atoms. Fig.~\ref{fig5:BetaF4} shows a storage time measurement of
$\rm{Cs}_2$$(v'=32-47)$ dimers together with a fit to
Eq.~\eqref{Eq:N2(t),F=4}. Several measurements at different atomic
densities yield fitted rate coefficients independent of atom
density as expected for atom-molecule collisions. Their average is
$\beta_{v'=32-47}^{F=4}=1.38\pm 0.04_{\rm{stat}}\pm
0.4_{\rm{sys}}\times 10^{-10}\unit{cm^3/s}$. This rate coefficient
is larger than $\beta_{v'=32-47}^{F=3}$, which can be attributed
to the additional loss channel of hyperfine-changing
collisions~\citep{Mudrich-collisions}.

%

The measured rate coefficients are quite close to the ones
predicted for $\rm {Na}-\rm {Na}_2$ and $\rm K-\rm K_2$
collisions~\citep{Quemener-NaNa2, Quemener-KK2} for low lying
vibrational levels in the respective $a^3\Sigma_u^+$ triplet
ground states and comparable to the value found in
Ref.~\citep{Mukaiyama-CollNaMol} for $\rm {Na}_2$ formed via a
Feshbach resonance. We do not, however, observe a strong increase
of the inelastic rate coefficient with the vibrational quantum
number, as predicted, e.g., for He-H$_2$
collisions~\cite{Balakrishnan-HeH2}. It is illustrative to compare
the measured inelastic rate coefficients to the inelastic s-wave
collision limit, $<\sigma^{(0)}v>\sim\sqrt{2\pi\hbar^4/(\mu^3 k_B
T)}$, where $\sigma^{(0)}=\pi/k^2$ is the s-wave limit of the
scattering cross section, $\hbar k$ the relative particle
momentum, $v$ the relative particle velocity, $\mu$ the reduced
mass and $<..>$ indicates an average over a Boltzmann
distribution. For collisions between Cs and $\rm{Cs}_2$ at
$T=60\unit{\mu K}$ we obtain $1.7\times 10^{-11}\unit{cm^3/s}$,
six times less than the measured rate coefficients. Since the
$l$'th partial wave at most contributes $(2l+1)\sigma^{(0)}$ to
the total scattering cross section it is clear that p-wave as well
as d-wave collisions play a significant role in the collisions.
This observation agrees with p-wave and d-wave barrier heights,
estimated as in Ref.~\citep{Cvitas-LiLi2}, of 16\unit{\mu K} and
83\unit{\mu K}, respectively.

In conclusion, we have presented a quantitative experimental study
of ultracold atom-molecule collisions with molecules in states
well below the dissociation limit. We have determined rate
coefficients for inelastic collisions between molecules in various
ro-vibrational states and Cs(F=3) atoms and in one case also
Cs(F=4) atoms at temperatures around 60\,$\mu$K. The inelastic
collisions are found to be governed by vibrational deexcitation
and show no dependence on the rotational quantum number. The rate
coefficients are about six times larger than the s-wave scattering
limit, showing that p-wave as well as d-wave collisions
contribute. In an improved setup containing a crossed dipole trap
providing higher atomic and thus molecular densities, it will be
possible to also study molecule-molecule collisions at ultralow
temperatures. Interesting future experiments will comprise
collisions and ultracold exchange reactions with mixed alkali
species in an optical trap \cite{mudrich2002:prl} and with
molecules in a single internal quantum state, particularly
molecules in $v=0$ of the singlet or triplet ground state, which
should have significantly longer trapping times due to the lack of
vibrationally inelastic
collisions~\citep{Sage-RbCs-v0,Meerakker-OHvibLifetime}. With the
development of high-resolution state-selective detection
techniques, studies on state-to-state molecular reaction dynamics
and quantum chemistry at ultralow temperatures seem to be within
reach.

\begin{acknowledgments}
This work is supported by the priority program SPP 1116 of the
Deutsche Forschungsgemeinschaft and by the EU Network `Cold
Molecules`.

\emph{Note added:} We recently became aware that similar work with
comparable results has been carried out simultaneously in the
group of P.~Pillet at the Laboratoire Aim\'e Cotton in Orsay,
France.
\end{acknowledgments}

\end{document}